\documentclass[aps,prl,preprint,floatfix,superscriptaddress]{revtex4-2}
\usepackage{hyperref,graphicx,bm}
\pdfoutput=1 

\begin{document}

\title{Theory of defect-induced crystal field perturbations in rare earth magnets}
\author{Christopher E.\ Patrick}
\email{christopher.patrick@materials.ox.ac.uk}
\author{Yixuan Huang} 
\affiliation{
Department of Materials, University of Oxford,
Parks Road,
Oxford OX1 3PH, UK}
\author{Laura H.\ Lewis} 
\affiliation{Department of Chemical Engineering, Northeastern University, Boston MA, USA}
\affiliation{Department of Mechanical and Industrial Engineering, Northeastern University, Boston MA, USA}
\author{Julie B.\ Staunton}
\affiliation{Department of Physics, University of Warwick, Gibbet Hill Road, Coventry CV4 7AL, UK}

\date{\today}

\begin{abstract}
We present a theory describing the single-ion anisotropy
of rare earth (RE) magnets in the presence of
point defects.
Taking the RE-lean 1:12 magnet class as a prototype,
we use first-principles calculations to show how the introduction of
Ti substitutions into SmFe\textsubscript{12} perturbs 
the crystal field, generating new coefficients due to the lower
symmetry of the RE environment.
We then demonstrate that these perturbations can be described
extremely efficiently using a screened point charge model.
We provide analytical expressions for the anisotropy energy
which can be straightforwardly implemented in atomistic spin
dynamics simulations, meaning that such simulations can be 
carried out for an arbitrary arrangement of point defects.
The significant crystal field perturbations calculated
here demonstrate that a sample which is single-phase
from a structural point of view can nonetheless have
a dramatically varying anisotropy profile at the atomistic
level if there is compositional disorder, 
which may influence 
localized magnetic objects like domain walls or skyrmions.
\end{abstract}

\maketitle 

The unique magnetic properties of
lanthanide elements originate from
their partially-filled 4$f$ shells~\cite{Elliottbook}.
Their
aspherical
charge distribution can yield a highly anisotropic
dependence of the total energy on the 
magnetic moment direction, i.e.\ a gigantic magnetocrystalline anisotropy (MCA),
which has been exploited with huge commercial success
in rare earth-transition metal (RE-TM) permanent magnets~\cite{Strnat1967}.
RE atoms also lie at the center of recent fundamental research into single molecule
magnets~\cite{Gould2022}, skyrmionics~\cite{Zhou2021, Zuo2021} and other
forms of topological magnetism~\cite{Yao2023}.
A quantitative description of RE MCA is provided by
crystal field (CF) theory~\cite{Newman1989}, where
the environment of each RE atom 
is described using a set of CF coefficients.
These are inserted via Stevens operators~\cite{Stevens1952}
into quantum or classical
Hamiltonians acting on the 4$f$ electrons~\cite{Kuzmin2008},
from which one can extract equilibrium properties through 
statistical mechanics~\cite{Yoshioka2022}, or carry out time-evolving
simulations within the framework of atomistic spin dynamics (ASD)~\cite{Evans2018}.
Further coarse-graining the atomistic results 
yields expressions for the MCA energy
for use in continuum micromagnetics~\cite{Fischbacher2018}.

A key target in permanent magnet development is 
a high maximum energy product $(BH)_{\mathrm{max}}$,
which in turn requires a high coercivity (resistance to demagnetization by external fields).
Although a large MCA is a prerequisite for a large, shape-independent coercivity~\cite{Brown1957},
RE magnets typically only achieve 5--35\% of their theoretical
limit~\cite{Nakamura2017}.
According to an influential analysis by 
Kronm\"uller~\cite{Kronmuller1987}, this performance gap
is due to inhomogeneities in the sample 
(e.g.\ misaligned or nonmagnetic grains) creating
magnetically soft regions susceptible to 
the nucleation of domains with reversed magnetization.
These grow
through domain wall motion and eventually
lead to destruction of the original magnetic order.

Kronm\"uller's analysis is a top-down approach, 
which posits that an inhomogeneity 
induces a variation in the anisotropy according to an assumed 
analytical profile.
However, computational magnetics is now focused on 
bottom-up simulations, whereby magnetization reversal and 
domain wall motion are treated as phenomena emerging from 
atomistic properties~\cite{Westmoreland2018}.
ASD simulations of homogeneous materials
are able to reproduce intrinsic quantities in impressive agreement
with experiment~\cite{Gong2019,Toga2018}.
However, to better understand the coercivity gap, it is
essential to develop a method of
incorporating inhomogeneities into these simulations~\cite{Vedmedenko2020}.

As an example, we consider how an ASD
simulation based on a classical spin Hamiltonian should 
be modified to include a point defect, the smallest possible 
inhomogeneity.
In ASD, each magnetic atom contributes to the total energy
based on the orientation of its magnetic moment
$\mathbf{\hat{e}} = (\sin\Theta\cos\phi, \sin\Theta\sin\phi, \cos\Theta)$~\cite{Evans2018}.
If a point defect is placed at a position $\mathbf{R_J}$
with respect to an RE atom, we derive the following 
additional contribution to the energy from the perturbed crystal field:
    \begin{equation}
    \Delta E_{\mathrm{CF}}(\mathbf{\hat{e}},\mathbf{R_J}) = \sum_{lm} \mathcal{A}_l
d^{0m}_l(\Theta)e^{im\phi} \Delta B_{lm}(\mathbf{R_J})
\label{eq.ASD}
\end{equation}
Both $d^{0m}_l(\Theta)$ and $\mathcal{A}_l$
are known quantities: $d^{0m}_l(\Theta)$ is a function
related to the associated Legendre polynomials~\cite{Edmonds,suppinfo}, and
$\mathcal{A}_l$ is an RE-dependent prefactor formed from
Stevens coefficients and the total angular momentum $J$~\cite{Sievers1982,Stevens1952}.
$l$ and $m$ are the usual angular quantum numbers.
Therefore, the only unknowns in equation~\ref{eq.ASD} are 
the $\Delta B_{lm}$ quantities,
describing the defect-induced change in each CF coefficient.
It is the calculation of these quantities, which are also required 
in a quantum treatment of the RE magnetism, which is
the subject of this Letter.
Specifically, we initially obtain $\Delta B_{lm}$ using first-principles 
density-functional theory (DFT), then use the calculations to parameterize
a highly efficient model suitable for deployment in large-scale ASD simulations.

Point defects, and their effects on the MCA, are particularly important in
the ``1:12'' class of RE-TM alloys~\cite{Tozman2021}.
These compounds
are under intense investigation
as ``rare earth-lean'' magnets, with their RE:TM atom ratio of 1:12
being more economical than 1:7 or 1:8.5
as required by Nd\textsubscript{2}Fe\textsubscript{14}B or 
Sm\textsubscript{2}Co\textsubscript{17}.
They are, however, fundamentally different to 
these established magnets, because they
have compositional disorder;
the host compound REFe\textsubscript{12} is
not stable in bulk, so it is necessary to 
introduce a transition metal M, like Ti, to form 
REFe\textsubscript{12-$x$}M\textsubscript{$x$}, where $x \sim 1$~\cite{Buschow1991}.
Although the M atoms prefer to sit at particular
sites in the unit cell, the substitution is probabilistic, so they can be viewed as
substitutional point defects embedded in an REFe\textsubscript{12} host.
Despite promising intrinsic properties, sufficiently high
coercivities have not yet been achieved,
leading naturally to the question of how this intrinsic disorder affects
1:12 performance.

We have therefore focused on SmFe\textsubscript{12-$x$}Ti\textsubscript{$x$},
an exemplar 1:12 alloy for which high-quality
single crystal data exists~\cite{Diop2020}.
This and closely-related 
materials have been subject
to numerous first-principles studies, examining
different TM substitutions~\cite{Harashima2016,
Harashima20152,Odkhuu2020,Matsumoto2020,Korner2016,
Landa2022,Benea2022,Fukazawa20192,Fukazawa2022,
Dirba2020,Butcher2017},
the effects of N-doping~\cite{Miyake2014,Delange2017}
or RE replacement~\cite{Bhandari2022}.
Recent studies have also carefully calculated
particular materials properties, comparing directly
to experiment~\cite{Herper2022,Yoshioka2022,Tozman2022}.
Our study emphasizes the \emph{local} 
influence of the substitutional atoms
and resulting implications  
across all RE magnets.

\begin{figure}
    \includegraphics{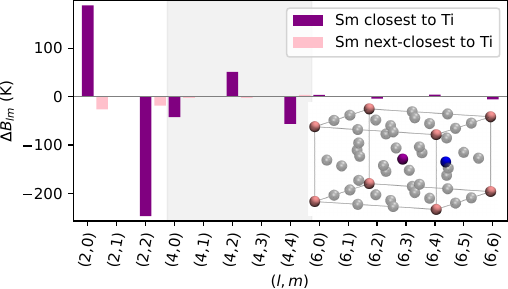}
    \caption{Calculated perturbations to the CF coefficients when 
    a single Ti atom (blue) is introduced at an 8$i$ site.  The Sm
    atoms at the 2$a$ sites are shown in purple and pink.  
    See Fig.~\ref{fig.2} for identification
    of the individual 8$i$, 8$j$, 8$f$ sites.  All
    CF coefficients are real, and $\Delta B_{lm}=\Delta B_{l-m}$.
    \label{fig.1}}
\end{figure}

We now describe our approach.
In CF theory, the potential $V({\bf r})$ 
experienced by the RE-4$f$ electrons is expanded in
terms of atom-centred functions, which here are
the complex spherical harmonics $Y_{lm}$,
i.e.\ $V({\bf r}) = \sum_{lm} V_{lm}(r) Y_{lm}({\bf \hat{r}})$.
The radial functions $V_{lm}(r)$ are then combined with the 4$f$ electron charge
density $n_{4f}(r)$ and integrated, to construct
the CF coefficients $B_{lm}$~\cite{Patrick20192}.
The anisotropy energy of the 4$f$ electrons is affected
only by CF coefficients with $l$ equal to 2, 4 or 6.
Furthermore, CF coefficients will be nonzero only if they
are compatible with the symmetry of the RE site.
SmFe\textsubscript{12-$x$}Ti\textsubscript{$x$} crystallizes
in the ThMn\textsubscript{12}-type structure 
shown in the inset of Fig~\ref{fig.1}.
The conventional unit cell is tetragonal and 
contains two formula units, with Sm atoms at the
corner and centre positions (Wyckoff label 2$a$) and transition metals 
occupying three distinct sublattices,
8$i$, 8$j$ and 8$f$.
In the pristine SmFe\textsubscript{12} host, these sublattices are
all occupied by Fe atoms.
This results in the 2$a$ sites
having $D_{4h}$ symmetry, whereby
the RE-4$f$ crystal field is specified completely
by the $(l,m)$ coefficients $(2,0)$, $(4,0)$, $(4,\pm4)$, 
$(6,0)$ and $(6,\pm4)$~\cite{BradleyCracknell,Kuzmin2008}.
We have calculated these within DFT
using the Y-analogue model~\cite{Patrick20192}, which we have
previously developed and applied to RE-Co, Nd-Fe-B and 
REFe\textsubscript{2}-type magnets~\cite{Patrick20193,Bouaziz2023,Patrick2020}.
Computational 
Details are provided at the end of this Letter.

Now we substitute one Fe atom with Ti at an 8$i$ site,
yielding
SmFe\textsubscript{11.5}Ti\textsubscript{0.5}.
As shown in the inset of Fig.~\ref{fig.1}, this 
substitution breaks the equivalence of the 2$a$ sites, with
the Sm atoms at the cell corners lying further away from the
Ti compared to the Sm atom located at the centre position.
Both 2$a$ sites have their symmetry lowered 
to $C_{2v}$.
Figure~\ref{fig.1} shows the effect on the crystal field
at the two different sites, plotting the change in
each coefficient $\Delta B_{lm}$.
We see that, as well as perturbing the existing nonzero
CF coefficients like $B_{20}$, the Ti substitution generates
new coefficients like $B_{22}$, due to the
lower symmetry.
The perturbations show two general features: first, that the largest
changes occur at the 2$a$ site closest to the 
substituted Ti, and second, that the largest changes
are in the $l=2$ coefficients.
The same behavior occurs when the Ti is
substituted at an 8$j$ or 8$f$ site~\cite{suppinfo}.

\begin{figure}
    \includegraphics{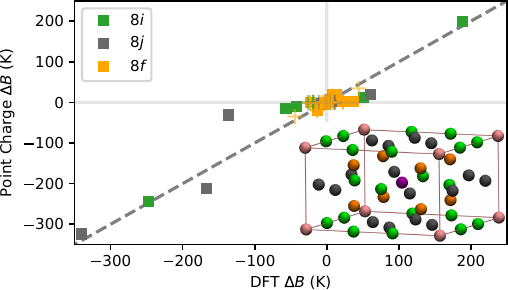}
    \caption{Perturbation to CF coefficients $\Delta B_{lm}$ calculated
    within DFT or using equation~\ref{eq.PCM}, when one Ti atom is placed
    in the conventional cell at an 8$i$, 8$j$ or 8$f$ site.  All $(l,m)$
    coefficients are shown for both 2$a$ sites.  + symbols are imaginary parts.  The inset shows the different sites
    in pristine SmFe\textsubscript{12}.
    \label{fig.2}}
\end{figure}

The strong $l$- and interatomic distance dependence of the CF perturbation
brings to mind the Laplace expansion of the Coulomb potential
and suggests that the Ti substitution might be efficiently modeled
as a screened point charge~\cite{Newman1989}.
We use this idea to derive 
an expression (in Hartree atomic units) 
for the change
in CF coefficient when a dopant is introduced at site $J$:
\begin{equation}
    \Delta B_{lm} = -\langle r^l\rangle
\left(\frac{4\pi}{2l+1}\right)^{\frac{1}{2}} 
\frac{\Delta Z_J}{R_J^{l+1}} \  Y^*_{lm}(\mathbf{ \hat{R}_J})
\label{eq.PCM}
\end{equation}
We have previously calculated 
$\langle r^l\rangle = \int r^{l+2}n_{4f}(r)dr$ with
self-interaction-corrected DFT~\cite{Patrick20192,Lueders2005} and found
it to be largely independent of the RE environment;
for Sm, $\langle r^l\rangle$
is 1.02, 2.46 and 10.25 for $l$ = 2, 4 and 6~\cite{suppinfo}.
$\Delta Z_J$ is the point charge associated with
the dopant, which depends on whether
site $J$ is 8$i$, 8$j$ or 8$f$.
To account for the short-ranged behavior shown in Fig.~\ref{fig.1},
we impose a cutoff radius $R_{\mathrm{c}}$ 
such that only defects with 
$R_J < R_{\mathrm{c}}$ contribute to $\Delta B_{lm}$.
Therefore, our model has four parameters, which 
we determine 
from a least-squares fit to the 
DFT calculations of $\Delta B_{lm}$,
(Fig.~\ref{fig.2}):
we find $\Delta Z_{8i} = 0.238$,
$\Delta Z_{8j} = 0.251$,
$\Delta Z_{8f} = 0.046$ and $R_{\mathrm{c}} = 4$~\AA.
The model captures the most significant changes in CF
coefficients, with errors in smaller $\Delta B_{lm}$ values of
order 50~K.

We note that the use of analytical models is hardly a new idea in
CF theory~\cite{Newman1989}, with Li and Cadogan's 
bonding charge model applied specifically to 1:12 compounds~\cite{Li1992}.
However, our current approach carries two particular advantages.
First,
we are parameterizing the model using direct DFT calculations of CF coefficients,
rather than having to extract them indirectly from experimental measurements~\cite{Diop2020,Tereshina2018}.
Second, we are only using the analytical model to describe the \emph{perturbation}
to the crystal field.
This both reduces the number of parameters in the model
and also increases the number of data points that can be used in the fitting
process, since the lower symmetry produces more nonzero CF coefficients.

\begin{figure*}
    \centering
    \includegraphics{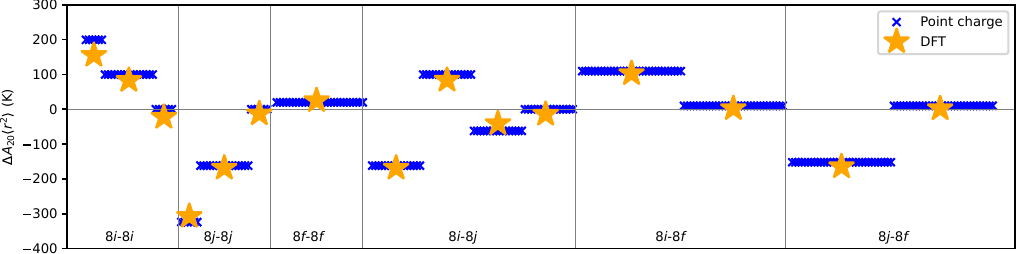}
    \caption{Perturbation to the $A_{20}\langle r^2\rangle$ CF coefficient when two Ti defects
    are introduced into the conventional unit cell at the indicated crystal sites,
    demonstrating excellent agreement between point charge (blue cross) or DFT (stars) calculations.
    \label{fig.3}}
\end{figure*}

Having established the model, we now consider its predictions.
Since equation~\ref{eq.PCM} is an electrostatic model, the principle of superposition
applies, such that introducing multiple substitutions has an additive 
effect on the CF coefficients.
At SmFe\textsubscript{11}Ti stoichiometry, two of the 24 TM sites 
in the conventional cell are occupied by Ti atoms, on average.
It is a simple matter to generate the 276 possible configurations (many of which are not unique),
and use equation~\ref{eq.PCM} to evaluate the perturbation to the CF.
The predicted effect on the $(2,0)$ coefficient is shown in Fig.~\ref{fig.3} as blue crosses.
Here,
the CF is quantified using the conventional $A_{20}\langle r^2\rangle $ coefficient,
which is equal to half of $B_{20}$~\cite{Newman1989,Patrick20192}.
The largest changes are predicted to occur when both Ti atoms are located either on 8$i$ or on 8$j$ sites, close
(within $R_{\mathrm{c}}$) to the RE atom.
Furthermore, due to opposing signs of $\Delta B_{20}$ for 8$i$ or 8$j$, there is a cancellation
effect for mixed 8$i$-8$j$ substitution.

To verify these predictions, we took a subset of the 276 configurations and calculated the CF coefficients
using DFT.
The resulting $\Delta A_{20}\langle r^2\rangle$ values are shown as stars in Fig.~\ref{fig.3}, 
and emphasize the 
excellent qualitative and quantitative agreement between the two methods.
Similarly good agreement is found for other coefficients, e.g.\ 
$\Delta A_{22}\langle r^2\rangle$ ~\cite{suppinfo}.
We stress the independence of the training set in Fig.~\ref{fig.2} and the test set in
Fig.~\ref{fig.3}, i.e.\ the data in Fig.~\ref{fig.3} was not used to refine the parameters.

Figure~\ref{fig.3} shows how the $A_{20}\langle r^2\rangle$ CF coefficient is 
affected by several hundreds of K when one or two Ti atoms are substituted close to the RE site.
The three DFT-calculated values of $\Delta A_{20}\langle r^2\rangle$ 
shown in Fig.~\ref{fig.3} for two 8$i$ substitutions
are +155, +83 and -23~K, corresponding to two, one and zero
Ti atoms lying close to the RE site, respectively.
To place these numbers into context,
calculations of $A_{20}\langle r^2\rangle$ for the REFe\textsubscript{12} host
find values between -50 and -200~K~\cite{Harashima2015,Delange2017,Yoshioka2020}, with our own calculations
giving -88~K or -136~K depending on whether experimental or DFT-optimized internal
co-ordinates are used.
Meanwhile, the latest experimental analysis of magnetization curves 
deduced a value of  $A_{20}\langle r^2\rangle$
of -110~K for SmFe\textsubscript{11}Ti~\cite{Diop2020}.
This agrees very well with the value obtained by 
combining $A_{20}\langle r^2\rangle$ for SmFe\textsubscript{12} 
from a detailed dynamical mean field
theory treatment of on-site correlation 
(-198~K)~\cite{Delange2017} with our calculated off-site electrostatic 
effect of Ti doping, averaged over configurations (+72~K).

According to our calculations, if two Ti atoms occupy 8$i$ sites
close to Sm, the perturbation to the CF will be so large that it will
almost destroy the single-ion anisotropy.
Therefore, it is crucial to know how the dopants are distributed.
We calculated the energies of all possible arrangements
of Ti atoms within the conventional cell of the analogous 
material YFe\textsubscript{11}Ti, and like previous
calculations and experiments on 1:12 systems~\cite{Ochirkhuyag2022,Matsumoto2020,Harashima20152,Yang1988},
found that the Ti atoms overwhelmingly prefer 8$i$ sites.
Importantly, we also found the most stable configuration has one Ti atom per (001) plane, which corresponds to both 
Sm sites having the same +83~K shift in $A_{20}\langle r^2\rangle$ shown in Fig.~\ref{fig.3}.
However, there are configurations which are higher in energy by only 6~meV/FU where alternate (001)
planes contain two or zero Ti atoms respectively~\cite{suppinfo}.
Such a structure would have an intrinsically textured anisotropy, where Sm sites in adjacent
(001) planes have alternating $A_{20}\langle r^2\rangle$ 
shifts of +155 and -23~K.
Given the closeness in energy of these configurations, it is highly likely that
synthesized SmFe\textsubscript{11}Ti samples will have Sm
atoms with a distribution of anisotropies, with  $A_{20}\langle r^2\rangle$ values  ranging
from $\sim$ -200 to -50~K.
Furthermore, any Ti-rich mesoscopic regions would have a significantly 
weakened (possibly even planar) anisotropy.

Although we have focused on the $(2,0)$ coefficient, Fig.~\ref{fig.1} clearly shows
how the $(2,\pm2)$ coefficient is also substantial.
It is important to realize that, assuming that the Ti dopants are uniformly distributed 
across a set of Wyckoff sites, the value of $\Delta B_{lm}$ \emph{averaged over 
all Sm sites} will be zero unless $m=0$ or $\pm4$.
Physically, this averaging process is equivalent to treating all of the Sm atoms
as a single entity---a magnetic sublattice--- with a single, common magnetization direction,
and is valid when comparing to measurements of macroscopic phenomena of bulk crystals,
e.g.\ measurements of magnetization like in Ref.~\cite{Diop2020}, which see an
averaged value of $\sim - 100$~K for $A_{20}\langle r^2\rangle$.
However, any noncollinear and/or localized magnetic texture,
such as a domain wall or a skyrmion~\cite{Zuo2021,Zhou2021}, 
will probe the anisotropy of only a subset of the RE atoms, leading
to an incomplete averaging.
Therefore, in simulating these phenomena, it is essential to consider
these additional CF coefficients,
even if they average to zero in the macroscopic limit.
We believe that there may be an interesting link between our work
and experiments carried out in the 1970s
investigating an ``intrinsic coercive force'
in ternary RETM\textsubscript{5} 
alloys like
SmCo\textsubscript{5-$x$}Ni\textsubscript{$x$}~\cite{Buschow1975,
Buschow1976}.
Intriguingly, it was found that even in samples containing only
a single phase, there existed a coercive field whose magnitude
depended on the concentration
of the substituted element.
It was pointed out that the Bloch walls separating domains
should only be a few atomic layers thick~\cite{Zijlstra1970,vandenBroek1971},
and it was argued that the dopants affected these through
modification of the exchange interaction.
However, the CF perturbation due to dopants was not 
considered, and it is important to 
revisit this phenomenon using our new approach.

We have focused our discussion on magnetocrystalline anisotropy, but it
is important also to quantify
the effect of Ti substitutions on the RE-TM exchange.
Using the DFT formulation of the disordered local moment
picture (DFT-DLM~\cite{Gyorffy1985,Patrick2022,Patrick20182,MendiveTapia2019,Daene2009}),
we calculated that the exchange fields at either of the two Sm sites
in SmFe\textsubscript{11.5}Ti\textsubscript{0.5}
are within 3 or 7\% of those in SmFe\textsubscript{12}, at 300~K.
We found a similarly small variation for
SmFe\textsubscript{11}Ti, even with the Ti 
defects clustered around only one Sm site~\cite{suppinfo}.
Based on these calculations, we conclude that the Ti dopants
do not appreciably modify the RE-TM coupling to Sm.

We have applied the point charge model to another
atomistic defect:
the RECo\textsubscript{5} 
``dumbbell'' 
whereby an RE atom is substituted
with a Co\textsubscript{2} pair~\cite{Kumar1988}.
Carrying out the same procedure as for
SmFe\textsubscript{12-$x$}Ti\textsubscript{$x$},
we find a
negative value of $\Delta Z_{\mathrm{Co_2}}$, -0.465,
and that the only RE sites affected 
by the presence of the Co\textsubscript{2} pair are the nearest neighbors
lying directly above or below it.
The negative value of $\Delta Z_{\mathrm{Co_2}}$ has a detrimental effect
on the anisotropy; pristine SmCo\textsubscript{5} has very
strong uniaxial anisotropy, with the prolate electron cloud
pointing along the $c$-axis~\cite{Patrick20193}.
Introducing the negative point charge along the same direction
reduces the magnitude of $A_{20}\langle r^2\rangle$.
This microscopic behavior is reflected in the bulk properties
of Sm\textsubscript{2}Co\textsubscript{17}, which can be
viewed as an ordered array of Co\textsubscript{2} defects
implanted in SmCo\textsubscript{5}~\cite{Kumar1988}, and
has a substantially reduced $A_{20}\langle r^2\rangle $
(50\%) compared to 
SmCo\textsubscript{5}~\cite{Patrick20192}.

As a summary, we set out our proposed scheme to carry out
ASD simulations of localized magnetic phenomena in the presence 
of point defects.
First, the Hamiltonian of the host material is constructed, taking
advantage of the widening literature of CF
coefficients and exchange constants for pristine systems.
Next, distributions of point defects are generated,
which may be totally disordered, partially or fully ordered;
the defect distributions might be built based on 
atom probe tomography data, particularly when
such measurements correlate with substantial variations
in coercivity~\cite{SepehriAmin2017}.
Equation~\ref{eq.PCM} is then used to calculate the CF 
perturbations at each RE atom, which are then fed
into equation~\ref{eq.ASD} to produce the Hamiltonian; an
example of this step is provided as Supplemental Information~\cite{suppinfo}.
Our DLM calculations indicate it is not necessary to modify the exchange
part of the Hamiltonian, and we also suggest that only the $l=2$ terms
need to be included in the CF perturbation.
Ideally, the $\Delta Z$ parameters should be obtained
by fitting to DFT calculations, but 
we stress their physical meaning
as being the effective charge of the defect.
Therefore, an order of magnitude estimate for $\Delta Z$ could be obtained 
from a relatively simple DFT calculation on the defective system, obtaining
the charge e.g.\ through a Bader analysis~\cite{Tang2009}.
From here, all of the sophisticated 
techniques developed for perfect systems~\cite{Gong2019,Toga2018}
may be applied.
The framework presented here thus opens a new avenue
in computational magnetics research, enabling the 
atomistic study of domain wall propagation and other localized
magnetic phenomena in systems with inhomogeneous
anisotropy.

\emph{Computational Details:}
CF coefficients were calculated within DFT using
the Y-analogue formalism as described in Ref.~\cite{Patrick20192},
using the \texttt{GPAW} code~\cite{Enkovaara2010}.
A plane-wave basis with cutoff energy of 800~eV and 6$\times$6$\times$6
reciprocal space sampling were used, and 
exchange and correlation effects were treated within
the local-spin-density approximation (LSDA)~\cite{Vosko1980}.
The structural parameters of the SmFe\textsubscript{12} cell
($a$, $c$ = 8.497, 4.687~\AA, $x_{8i}$ = 0.359, $x_{8j}$ = 0.270)
were taken from previous DFT 
(generalized-gradient approximation) calculations~\cite{Harashima2015}
and were used throughout the study, except for the total energy comparison
and indicated CF calculations,
where the internal co-ordinates were allowed to relax.
The DFT CF coefficients were averaged over spin directions
before fitting to the point charge model.

\begin{acknowledgments}
We thank A.\ Gabay and G.\ Hadjipanayis for useful
discussions.
This work
was supported in part
by a Royal Society Research Grant No. RGS\textbackslash R1\textbackslash201151
and by the U.S. Department of Energy,
Office of Basic Energy Sciences under Award Number DE
SC0022168.
Y.\ H.\ acknowledges funding from the Ironmongers' Foundation.
We acknowledge the University of Oxford 
Advanced Research Computing (ARC) facility in carrying out
this work~\cite{ARC}.
Figures were rendered using VMD~\cite{VMD}.
\end{acknowledgments}

%\bibliography{papers}

\end{document}